# Sound Absorption and Transmission Loss Properties of Open-Celled Aluminum Foams with Stepwise Relative Density Gradients

Amulya Lomte [a], Bhisham Sharma [a✉], Mary Drouin [b], Denver Schaffarzick [c]

[a] Mechanics and Dynamics Laboratory, Department of Aerospace Engineering, Wichita State University, Wichita, KS, 67260, USA
[b] Spirit Aerosystems, Wichita, KS, 67210, USA
[c] ERG Materials and Aerospace Corporation, Oakland, CA, 94608, USA

✉ Corresponding author:
Bhisham Sharma, Mechanics and Dynamics Laboratory, Department of Aerospace Engineering, Wichita State University, Wichita, KS 67270, USA. Email: bhisham.sharma@wichita.edu

**Abstract**

We investigate the acoustical properties of uncompressed and compressed open-celled aluminum metal foams fabricated using a directional solidification foaming process. We compressed the fabricated foams using a hydraulic press to different compression ratios and characterized the effect of compression on the cellular microstructure using microtomography and scanning electron microscopy. The static airflow resistances of the samples are measured and related to the observed microstructural changes. We measured the normal incidence acoustical properties using two- and four-microphone impedance tube methods and show that the compression substantially improves their sound absorption and transmission loss performance. We then stack individual disks with different compression ratios to create various stepwise relative density gradient configurations and show that stepwise gradients provide a significant improvement in properties as compared to the uncompressed sample. The effect of increasing and decreasing relative density gradients on the overall absorption and transmission loss behavior is characterized. Finally, we use an experimentally informed and validated transfer matrix method to predict the effect of various layer thicknesses and stacking sequences on the global acoustical properties. Our results show that open-celled metal foams with stepwise relative density gradients can be designed to provide tailored acoustic absorption performance while reducing the overall weight of the noise reduction package.

## 1. Introduction

First hypothesized by Alexandre De Meller in his 1925 patent application [1], metal foams have attracted increasingly growing attention since the mid-1990s due to a combination of factors: increasing fuel prices and global environmental consciousness motivated transportation system manufacturers to reduce component weights "anywhere we can" [2, 3]; stricter passenger safety regulations necessitated the development of structures with higher energy absorption capabilities [4]; and the publication of Gibson and Ashby's book titled "Cellular Solids" [5]—providing property charts and a state-of-art assessment of cellular metals—brought familiarity and ignited interest among engineers [6]. Metal foams provide unique property combinations that primarily depend on their base material, relative density, and cellular topology. Their high stiffness-to-weight ratios, exceptional strain energy absorption characteristics, and high thermal conductivity, as compared to their monolithic counterparts, make them ideal for lightweight structural applications. A detailed survey of their physical properties and potential applications is available in Refs. [5-9]; more recent reviews of their various mechanical and functional properties are available in Refs. [10, 11] .

The acoustic properties of metal foams depend on their cellular morphology. Closed-cell foams are poor absorbers since they predominantly reflect the incident sound waves. On the other hand, open-celled foams allow the incident waves to travel through their cells and dissipate the incident acoustic energy via frequency-dependent visco-inertial and thermal loss mechanisms—making them effective sound absorbers [12]. Though metal foams are not as acoustically efficient as traditional polymeric bulk absorbers of equivalent weight—stabilization issues limit the feasible reduction in individual cell sizes [13] which limits the energy conversion processes necessary for high absorption—the additional benefit of having a metallic cellular structure makes them ideal for certain applications. Metal foams are more weather durable and can provide the structural stiffness necessary for load-bearing applications [7]. Depending on their base material, metal foams can provide electromagnetic shielding [14], corrosion [15] and flame resistance [16], and remain stable at high temperatures and pressures [5, 7]. Thus, unlike polymeric foams, they can be implemented in harsh operational environments that can benefit from acoustical treatment such as excimer laser discharge chambers [17] and acoustic liners placed directly over the turbofan rotor for aircraft noise reduction [18].

Given their apparent multifunctional benefits, various researchers have investigated their acoustic properties with an emphasis on their sound absorption characteristics. Lu et al. [19] studied closed-cell aluminum foams (Al-foams) and showed that their sound absorption performance can be improved by cracking open the cells using either compression, rolling, or by drilling holes through every two or three cells. In effect, this converts them into open-celled foams and increases the thermoviscous losses. Of the three methods, hole drilling provided the highest enhancement in the sound absorption coefficient. Similar results were obtained by Fu-sheng et al. [20], who attributed the improvement in the absorption performance to the increase in flow resistance. Wang et al. [21] used a point-matching based semi-analytical formulation to study the effects of cellular morphology and cavity depth on the acoustic properties of hexagonal and random Al-foams. They observed that for a given porosity, while the cell shape has a minimal effect on absorption, a cell size on the order of 0.1 mm appears to provide optimum absorption for practical combinations of sample thickness, cavity depth, and porosity. Lu et al. [22] further analyzed the absorption behavior of semi-closed-cell Al-foams and showed that the absorption increases with decreasing pore sizes and an absorption coefficient larger than 0.8 could be achieved in the 800–2000 Hz frequency range. Later, similar results were obtained by Han et al. [23], who showed that, in accordance with earlier results by Fu-sheng et al., smaller pore sizes result in better sound absorption due to the increase in flow resistance. Similarly, Bai et al. [24] studied the effect of compression on the absorption performance of nickel-iron alloy foams and found that higher compression ratios result in higher sound absorption. Spurred by these results, Sutliff et al. [18, 25, 26] and Jones et al. [27] investigated the potential use of various open-celled metal foams as acoustic liners placed directly over an aircraft engine turbofan rotor. They achieved nearly 3 dB inlet and aft noise reduction during low-speed fan rig tests and up to 5 dB sound power level attenuation during tests conducted using a Williams International FJ44-3D engine. Later, Xu and Mao [28, 29] showed that open-celled metal foams are highly effective at suppressing tonal as well as broadband noise components of centrifugal fans. Liu et al. [30] investigated the possibility of reducing aerodynamic noise using open-celled metal foams and found that foams with high porosity reduce noise significantly by modifying the flow, adjusting the vortex shedding frequency and regularizing the wake. To avoid the high flow resistance penalty associated with smaller pore size, Ke et al. [31] fabricated open-celled Al-foams with graded pore sizes and showed that such foams provide better absorption properties than foams with uniform pore sizes. More recently, Yang et al. [32] fabricated and compressed open-celled copper foams and showed that a four-layer gradient compressed foam configuration can provide significantly improved absorption. To the best of our knowledge, no systematic studies on the transmission loss behavior of open-celled metal foams currently exist in the published literature.

Evidently, reducing the effective pore size, by either directly fabricating such foams or by compressing the prefabricated foams, improves the sound absorption behavior of metal foams. However, this reduction in pore size is typically accompanied by an undesired increase in the airflow resistance and relative density. Here, we systematically study the acoustic properties of uniformly compressed open-celled Al-foams and investigate the possibility of improving their overall absorption and transmission loss performance over a specified thickness by stacking individual foams with different compression ratios into a layered configuration—effectively creating a stepwise, spatial, relative density (or porosity) gradient. First, we study the effect of compression on the foam's cellular morphology, relative density, and airflow resistance. Then, we experimentally study their normal incidence acoustical properties using impedance tube testing. We focus on the effect of compression and various stepwise relative density gradients on their absorption and transmission loss performance. Finally, using an experimentally informed and validated transfer matrix method, we study the effect of various layer configurations on the absorption and transmission loss behavior, and investigate the possibility of improving these properties over a wide frequency range without incurring an excessive weight penalty. The next section describes the experimental method and introduces the transfer matrix method, followed by a discussion of the obtained results.

## 2. Methods

### 2.1. Sample fabrication

In this study, we use open-celled Al-foams sold by ERG Materials and Aerospace Corporation under the commercial name: Duocel. The parameters for the proprietary foaming process are tuned to foam Al6061 billets with a base configuration of 7% relative density and 40 pores per inch (PPI). The compressed foams are then fabricated by hydraulically pressing previously foamed billets, with compression rates chosen to ensure uniform through-thickness compression. We maintain the final thickness of all individual samples as 12.7 mm (0.5 inch); hence, the initial thickness of the sample determines the compression ratio—defined as the ratio of the initial (pre-compression) and final (post-compression) sample thicknesses. We study the effect of four different compression ratios: 1, 2, 3, and 4, where a compression ratio of 1 indicates an uncompressed sample while a compression ratio of 4 indicates that the initial thickness of the sample was four times its final thickness. Representative examples of the uncompressed and compressed samples are shown in Fig. 1.

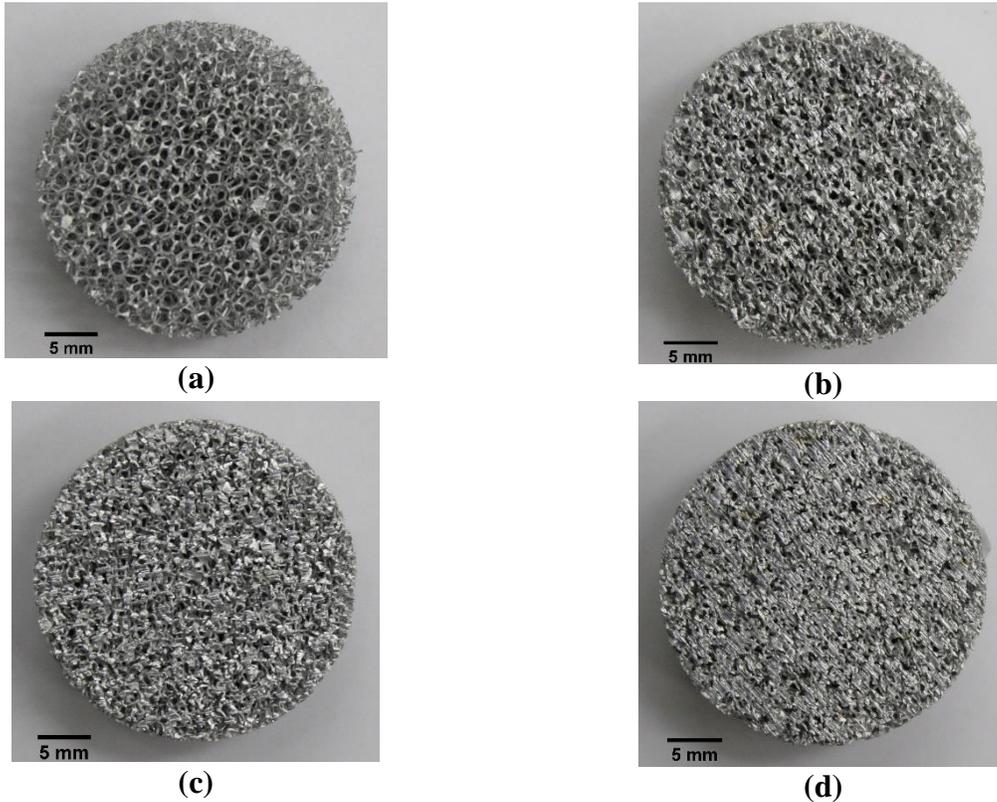

**Fig. 1.** Representative examples of the **(a)** uncompressed and **(b)**, **(c)**, **(d)** compressed foams with compression ratio of 2, 3, and 4, respectively.

For the impedance tube testing, all 12.7 mm thick foam plates are turned into 30 mm and 60 mm diameter disks. We then stack individual disks within the impedance tube sample holder to achieve either uniform 50.8 mm (2 inches) thick samples by stacking four disks with the same compression ratio, or a stepwise through-thickness relative density gradient by stacking four disks of varying compression ratios. Note that the total thickness of all configurations, uniform or varying, is maintained at 50.8 mm for all impedance tube tests reported here, unless specified otherwise. To identify the different layered configurations, we use the following nomenclature: letters A, B, C, and D are used to represent Al-foams with a compression ratio of 1, 2, 3, and 4, respectively. Since the total sample thickness is achieved by stacking individual 12.7 mm thick disks, the numbers 1, 2, and 3 are then added to the respective letter to clarify the total thickness of that specific layer within a given configuration—i.e., 1 signifies a layer thickness of 12.7 mm, 2 signifies a total layer thickness of 25.4 mm, and 3 signifies a total layer thickness of 50.8 mm. For example, A2 represents a uniform configuration of Al-foam with a compression ratio of 1 (i.e., uncompressed) with a total thickness of 25.4 mm. As an example of a layered configuration, B1_C1_D2 represents a configuration where the first layer is an Al-foam with a compression ratio of 2 (B) of thickness 12.7 mm (1), the second layer is an Al-foam with a compression ratio of 3 (C) of thickness 12.7 mm (1), and the third layer is an Al-foam with a compression ratio of 4 (D) of thickness 25.4 mm (2). As a rule of thumb, the sound waves are always incident on the first layer in the sequence.

## 2.2. *Normal incidence impedance tube testing*

The acoustic properties of all sample configurations are measured using two-and four-microphone normal incidence impedance tube methods, as outlined in ASTM E1050 [33] and E2611 [34], respectively. The two-microphone configuration is used to measure the sound absorption and surface impedances, and the four-microphone configuration is used to measure the sound transmission loss and characteristic impedances. Fig. 2 shows the schematics of both test setups. In both setups, a power amplifier drives a loudspeaker placed at the left end of the tube to generate broadband white noise with frequency content from DC to 10 kHz, and the pressure measurements are taken using ¼ inch PCB model 130F22 ICP array microphones with integral preamplifiers. We perform two measurements for each layered configuration to get the plane wave, normal incidence acoustic properties over the 400–5500 Hz frequency range: the first set of measurements, conducted using a 60 mm diameter tube, provides data from 400 Hz to 2500 Hz; the second set of measurements, conducted using a 30 mm diameter tube, provides data from 600 Hz to 5500 Hz. The datasets are then combined to obtain the results over the entire frequency range.

For the two-microphone measurements, the samples are placed in an acoustically rigid holder, with the foam surface kept flush against the rigid backing surface for all measurements. To ensure plane wave conditions, microphone spacings, $s$, of 45 mm and 22.5 mm are used to measure the pressures within the 60 mm and 30 mm tube, respectively. For these configurations, the distance, $l$, between the sample incident surface and the reference microphone—the microphone closest to the sample—is maintained at 35 mm and 15 mm, respectively. We measure the transfer function, $H_{12}$, between the two microphones using a LMS SCADAS SX data acquisition system in conjunction with Siemens Testlab software. The sound reflection coefficient, $R$, is calculated using the measured transfer function as:

$$R = \frac{H_{12} - e^{-jks}}{e^{jks} - H_{12}} e^{j2k(l+s)} \tag{1}$$

where $k$ is the wave number and j is the complex number equal to $\sqrt{-1}$. The sound absorption coefficient, $\alpha$, and the surface impedance, $Z$, are then calculated as:

$$\alpha = 1 - |R|^2 \tag{2}$$

$$Z = \frac{1+R}{1-R} \tag{3}$$

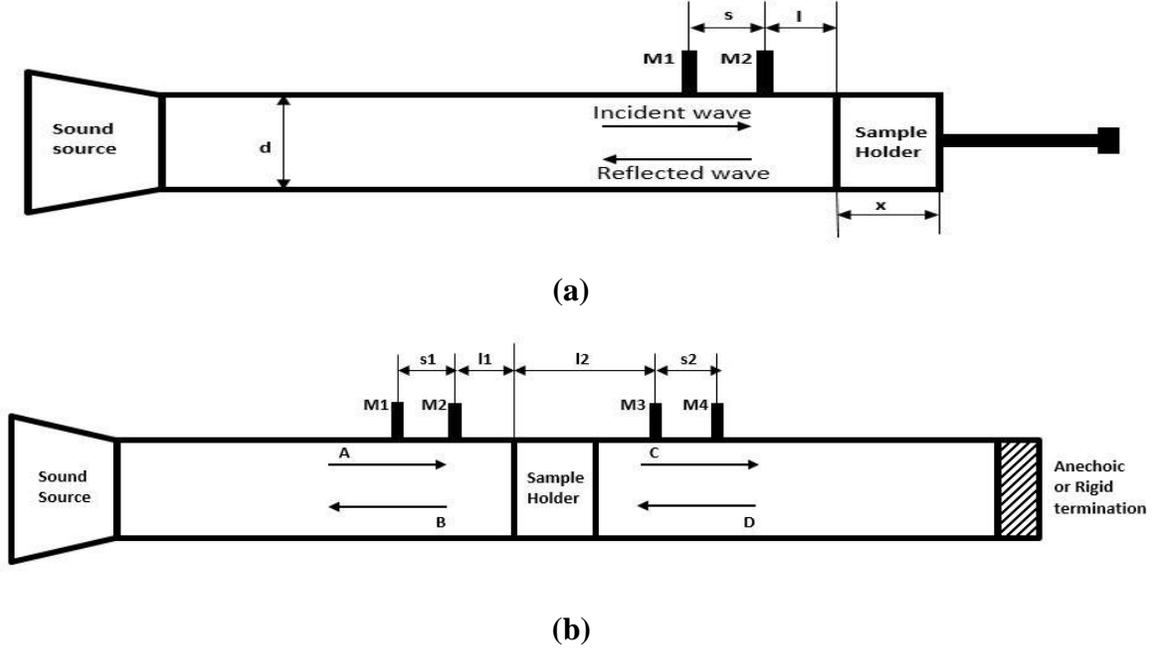

**(a)**

**(b)**

**Fig. 2.** Schematic of the **(a)** two-microphone and **(b)** four-microphone normal incidence impedance tube setups used to measure the acoustic absorption and transmission properties, respectively.

For the four-microphone measurements, the samples are placed between two sets of upstream and downstream microphones and each measurement is conducted using anechoic and rigid termination conditions, as detailed in Ref. [34]. The microphones are separated by 45 mm and 22.5 mm for tubes with a diameter of 60 mm and 30 mm, respectively. $l_1$ and $l_2$ represent the distances from the front surface of the sample to the nearest microphones located upstream and downstream of the sample. For 60 mm and 30 mm tubes, $l_1$ is 35 mm and 15 mm, respectively, while $l_2$ is maintained at 100 mm for both setups. The measured transfer functions are then used to calculate the transfer matrix elements, from which the transmission coefficient, $T_a$, sound transmission loss, TL, complex wave number, $k_p$, and characteristic impedance, $\zeta$, are calculated as:

$$T_a = \frac{2e^{jkd}}{T_{11} + \left(\frac{T_{12}}{\rho_0 c}\right) + \rho_0 c T_{21} + T_{22}} \tag{4}$$

$$\text{TL} = 20 log_{10} \frac{1}{|T_a|} \quad (5)$$

$$k_p = \frac{1}{d} \cos^{-1} T_{11} \quad (6)$$

$$\zeta = \sqrt{\frac{T_{12}}{T_{21}}} \quad (7)$$

where $\rho_0$ is the density of air, $c$ is the speed of sound in air, and $d$ is the sample thickness.

*2.3. Static airflow resistance measurement*

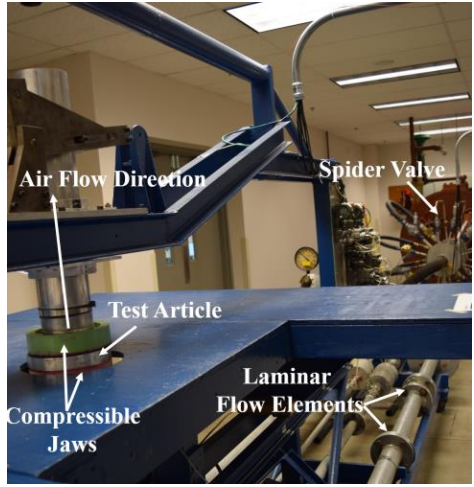

**Fig. 3.** Test setup used to measure the airflow resistivity of foam samples.

The static airflow resistance measurements are conducted using the flow resistance test setup shown in Fig. 3. The custom test setup comprises four laminar flow elements, a spider valve configuration, a thermocouple, and eight pressure transducers for the generation and accurate measurement of flow velocities. The sample is held in a holding fixture and the pressure drop across it is determined using the pressure, flow velocity, and measurement area as prescribed by the ASTM C522-03 test standard [35]. For each compression ratio, we tested two samples each by incrementally increasing the flow velocity from 0.1 m/s till 2 m/s. The measured flow resistances are then plotted as a function of the flow velocity and the static airflow resistance is calculated as the y-intercept of the flow resistance curve.

*2.4. Transfer matrix modeling*

The Biot theory [36, 37] provides an analytical framework for modeling sound propagation in a fluid saturated, isotropic, porous media. According to this theory, the acoustic behavior of a general poro-elastic medium is described by accounting for three propagating waves: one compression and one shear wave in the elastic skeletal structure and one acoustic compression wave in the fluid filled pores. If the medium primarily behaves as an elastic (solid) material, then the analysis can be simplified to only include the elastic compression and shear waves. The analysis of materials with either negligible (limp) or extremely high (rigid) bulk frame elasticity can be further simplified by modeling them as an equivalent fluid medium with only one propagating compression wave.

For cellular porous materials, the appropriate model choice depends on the stiffness of the skeletal frame and the frequency range of interest. Given their relatively high frame stiffness and extremely low estimated phase decoupling frequency [38], open-celled Al-foams can be modeled as rigid porous media with a motionless frame. The acoustic pressure and particle velocity at the incident and rear surfaces of the equivalent fluid medium are then related via a two-dimensional, square transfer matrix, **T**, given as:

$$\mathbf{T} = \begin{bmatrix} \cos(kd) & j\frac{\omega\rho}{k}\sin(kd) \\ j\frac{k}{\omega\rho}\sin(kd) & \cos(kd) \end{bmatrix} \quad (8)$$

where $\rho$ is the complex density of the equivalent fluid medium, $d$ is the layer thickness, and $\omega$ and $k$ are the angular frequency and wave number of the propagating acoustic wave. This transfer matrix modeling method—originally introduced by Mason [39] to study the combination of acoustical elements for efficiently designing acoustic filters—provides a convenient analytical approach for modeling layered acoustical media. In this method, to model porous multilayered configurations, appropriate transfer matrices describing each individual layer are first formulated. These individual transfer matrices may be directly measured or can be formulated using semi-empirical theories relating the material's microstructural details to its bulk properties, such as the frequently used Johnson-Champoux-Allard model [12, 40]. Then, the global transfer matrix, $\mathbf{T^G}$, describing the acoustical behavior of the layered configuration is calculated as:

$$\mathbf{T^G} = [\mathbf{T^1}][\mathbf{T^2}][\mathbf{T^3}]\ldots[\mathbf{T^i}] \quad (9)$$

where $\mathbf{T^i}$ represents the transfer matrix of the i[th] layer. Note that layers modeled as general poro-elastic materials require a six-dimensional transfer matrix and those modeled as solid elastic materials require a four-dimensional transfer matrix. The appropriate procedures for coupling such ill-dimensioned acoustical layers are available in Ref [41].

Here, we model the compressed and uncompressed Al-foam layers as rigid acoustical materials and get their individual transfer matrices using the four-microphone impedance tube measurements described in Section 2.2. We then calculate the global transfer matrix for each layered configuration using Eq. (9) and by assuming continuity of the acoustic pressure and normal particle velocities at the layer interfaces. The resulting global transfer matrix is then used to calculate the acoustic properties of various layered configurations and investigate the effect of different layer sequences and thicknesses on the overall normal incidence sound absorption and transmission loss performance.

### 3. Results and discussion

*3.1. Effect of compression on cellular morphology and flow resistivity*

Compressing the Al-foam billets changes their relative densities. Table 1 summarizes the measured weights, relative densities, and porosities of the 12.7 mm disks used in this study. The given weights are the averaged values of four individual disks and the relative densities are calculated assuming the density of the base aluminum as 2,710 kg/m$^3$. As expected, the relative density of the compressed samples is higher than the uncompressed sample, with samples of compression ratio 4 approximately 4.61 times heavier than the uncompressed samples.

**Table 1.** Comparison of the mass properties of compressed and uncompressed Al-foam samples.

| Compression Ratio | Measured Weight (g) | Relative Density (%) | Porosity (%) | Relative Density Ratio (with reference to uncompressed Al-foam) |
|---|---|---|---|---|
| 1 | 1.72 | 7.09 | 92.9 | 1 |
| 2 | 4.03 | 16.61 | 83.38 | 2.34 |
| 3 | 5.98 | 24.60 | 75.39 | 3.46 |
| 4 | 7.96 | 32.76 | 67.23 | 4.61 |

Importantly, the applied compression can potentially alter the geometric parameters of direct relevance to the foam's acoustic behavior—the pore structure and distribution. Here, we use x-ray microtomography and scanning electron micrography (SEM) to understand and quantify these morphological changes. Microtomography images are obtained using a North Star Imaging (NSI) X3000 3D x-ray computed tomography system with a 225 kV microfocus tube. The scans provide a 3D volume resolution of 1403 x 628 x 1355 voxels over the 12.7 mm disks, resulting in a focal spot size of 24.689 µm. The obtained images are first processed using a proprietary software (efX-ct) provided by NSI and then further analyzed using Image J. Here, for representational clarity, we export the volume to GOM Inspect and utilize its native capabilities to obtain the cross-sectional images shown in Figs. 4 and 5. The SEM images are obtained using a Phenom Pharos Desktop SEM system using an acceleration voltage of 10 kV. The SEM is equipped with a back scatter electron detector (BSE) and a secondary electron detector (SED). BSE and SED images obtained from multiple locations from each sample are exported and further analyzed using Image J. Considering their greater clarity, only the SED images are shown here in Figs. 6(a-f).

Figs. 4 and 5 show the microtomography images of an uncompressed sample and a sample with compression ratio of 4. In addition to the isometric view of the entire scanned samples, two cross-sectional slices along the thickness of the samples are shown for clarity; a 1 mm thick slice is shown for the uncompressed sample and a 0.5 mm thick slice is shown for the compressed sample. The uncompressed foam shows a relatively uniform pore structure with no visible morphological differences between the sample core and surface regions. While some cells break, the majority are intact with the individual cell geometries resembling a typical Kelvin cell structure [42] with varying amounts of distortion. The cell structure is more clearly visible in an SEM image of the uncompressed foam, as shown in Fig. 6(a). On average, the length of the struts ranges from 500 μm to 1 mm. The cross-sectional shape of the struts is triangular with round edges, with side widths observed to consistently range between 180 to 220 μm. The cellular microstructure shows a distinct lack of membranes observed by researchers studying other similar foams.

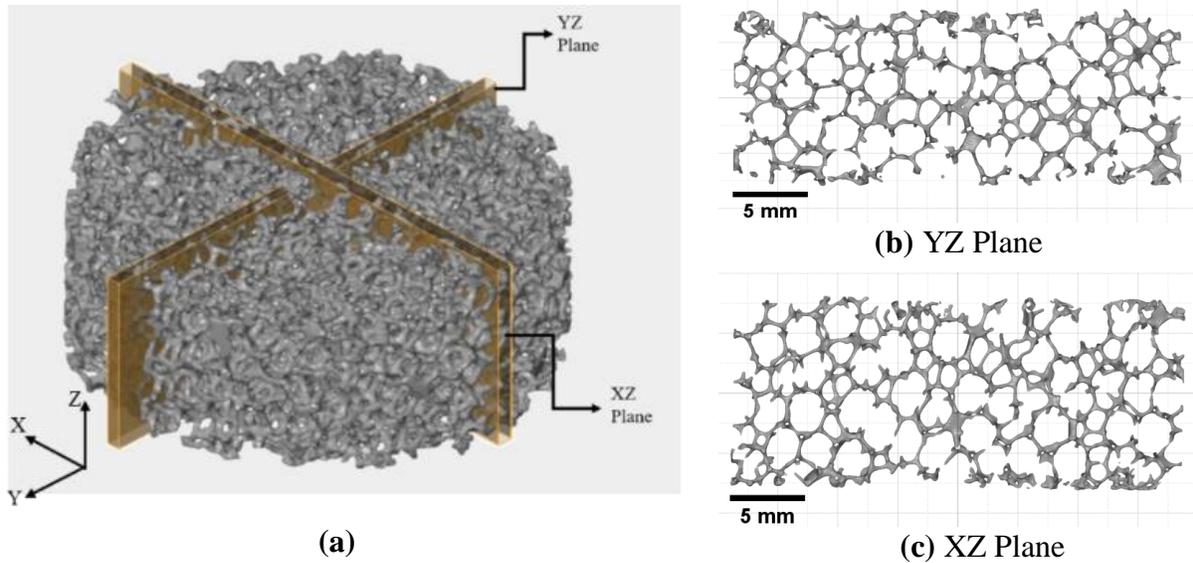

**Fig. 4.** Cellular morphology of uncompressed Al-foam visualized using microtomography imaging. (a) shows the location of the cross-sectional slices shown in (b) and (c). The shown slices are both 1 mm thick.

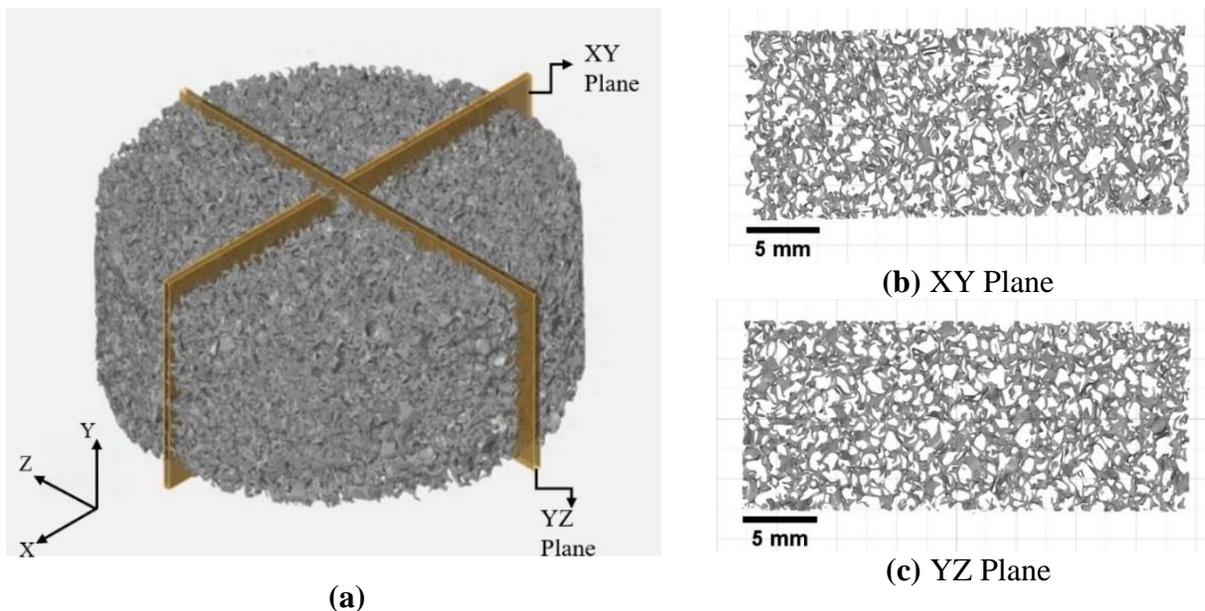

**Fig. 5.** Cellular morphology of Al-foam with compression ratio of 4, visualized using microtomography imaging. (a) shows the location of the cross-sectional slices shown in (b) and (c). The shown slices are both 0.5 mm thick.

Fig. 5 shows the effect of compression on the cellular morphology of the Al-foam. The applied compression distorts the cell structure throughout the sample, with no observable differences between the core and the upper and bottom surface layers. This agrees with previous work by Lu et al. [19] who observed that while the use of a roller to compress closed cell foams results in a core-shell pattern with greater crushing along the surface than the core, the direct application of compression, as done in this study, causes uniform distortion through the sample thickness. Overall, the application of compression reduces the open pore dimensions; the reduction is greater for samples with higher compression ratios.

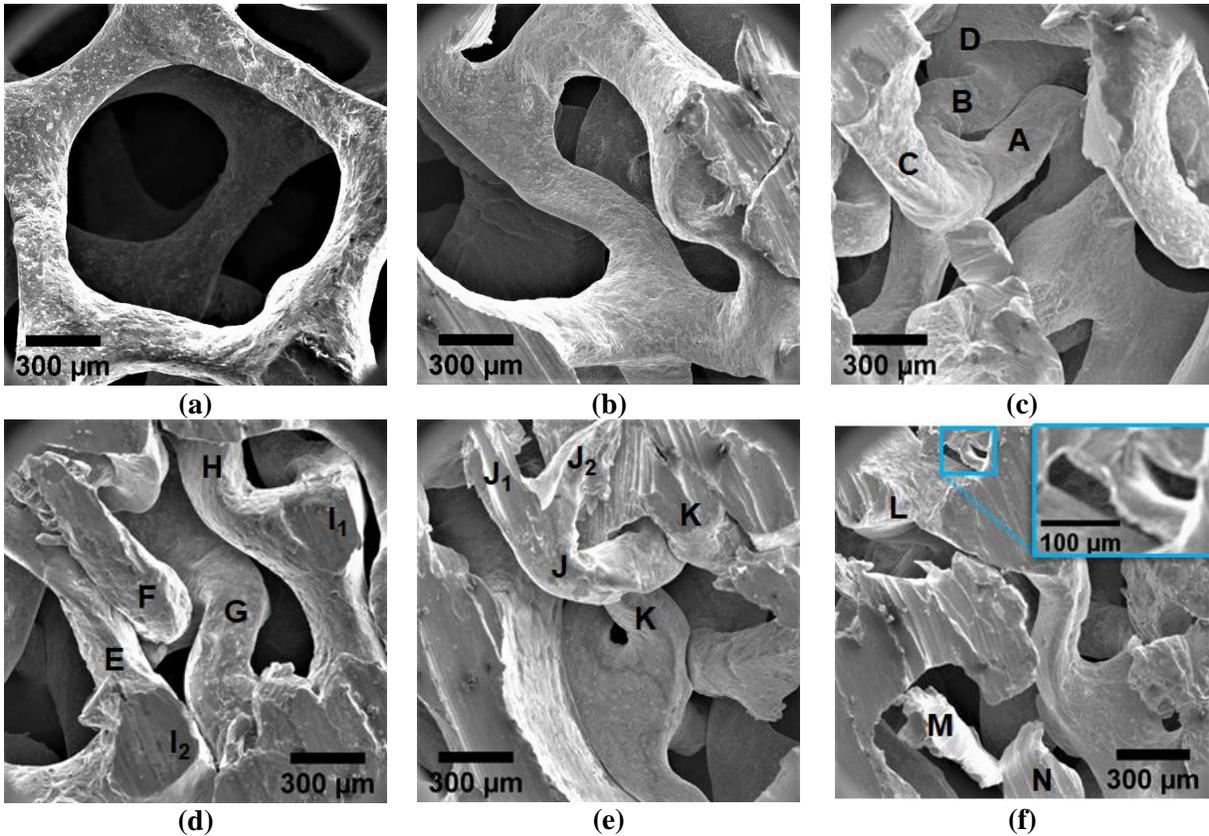

**Fig. 6.** SEM images of Al-foam samples with compression ratios of **(a)** 1 (uncompressed); **(b)** 2; **(c)** 3; and **(d-f)** 4. All images shown here are obtained using the secondary electron detector.

SEM images shown in Figs. 6(b-f) reveal the nature of distortion and its dependence on the compression ratio. All SEM images are taken normal to the surface where the compression was applied. Fig. 6(b) shows a representative image of the cell structure for the sample with a compression ratio of 2. The compression causes the individual struts to rotate and bend—the amount of rotation and bending in each strut is dictated by its original orientation vis-à-vis the loading direction and its length relative to the connected struts forming the individual three-dimensional cell. As the compression ratio increases to 3 (Fig. 6(c)), the bending and rotations increase, and some struts converge towards each other. In the shown image, the rotation about their connecting node causes struts A, B, and C to orient almost along a single plane, with struts A and B nearly touching each other. The high degree of torsion is evidenced by the relative orientations of strut B and D. Thus, the applied compression reduces the effective pore sizes and creates inter-strut crevices, which may cause additional acoustic energy losses through thermoviscous effects.

Figs. 6(d-f) show representative SEM images of a sample with compression ratio of 4. The increased compression results in densification and some struts—observe struts E, F, G, and H in Fig. 6(d)—compress against each other. While we observed little to no strut breakages in samples with lower compression ratios, samples with compression ratio of 4 show clear evidence of strut damage and breakages. Locations $I_1$ and $I_2$ in Fig. 6(d) show fracture mirrors on nodal surfaces, indicating an intergranular brittle fracture of struts because of excessive stresses caused by the bending moments [43]. Fig. 6(e) shows further evidence of densification: struts K and J, originating at neighboring nodes, interlock with each other and create additional narrow inter-strut crevices. Interestingly, strut J splits into two parts, $J_1$ and $J_2$, creating a sharp fracture surface between the two. Onck et al. [44] previously observed a similar damage mechanism, labeled "unzipping", during the *in situ* fracture study of open-celled nickel chromium alloy foams with cell structure like the foams studied here. This unzipping occurs due to preexisting imperfections in the hollow struts which result in fracture under the excessive shear forces from the applied bending and torsion. This creates additional surfaces and pores in highly compressed foams, potentially increasing the thermoviscous energy dissipation. Fig. 6(f) shows another example of an unzipped strut which creates a sharp pore with a width of approximately 20 μm, as seen in the zoomed inset. Location L shows further evidence of strut fracture, where the cupped surface and wake hackles indicate occurrence of a ductile fracture because of excessive tensile stresses generated in struts perpendicular to the loading direction [43]. Fig. 6(f) also shows broken strut pieces, M and N, stuck within the foam. These broken pieces further constrict open pores and increase the overall flow resistance of the foam.

Fig. 7 shows the measured airflow resistance as a function of flow velocity for the uncompressed and compressed samples. We tested two samples of each compression ratio to ensure repeatability. Then, the static airflow resistivity values, provided in Table 2, are calculated for each sample using its measured flow resistance. As expected, the uncompressed samples provided significantly low resistivity, indicating poor low frequency acoustic absorption capabilities. The resistivity increases with increasing compression ratio—samples with compression ratios 2, 3, and 4 are approximately 3, 7.5, and 20 times as resistive as the uncompressed samples, respectively. The significant increase in resistivity between the samples with compression ratios 3 and 4 again shows that the sample has started densifying and further compression would not provide substantial acoustic performance benefits. This agrees with the observation previously made using the SEM images. It is important to note here that though the resistivity of the compressed samples is higher than the uncompressed sample, the values are still lower than typical resistivity values (>10,000 MKS Rayls/m) of traditional acoustic materials like fiberglass, rockwool, and polyurethane foams.

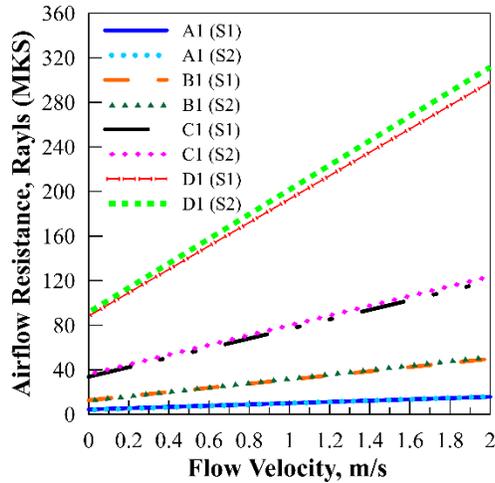

**Fig. 7.** Measured airflow resistance as a function of the flow velocity. The static airflow resistivity is obtained by dividing the y-intercept of the measured curves with the test sample thickness (12.7 mm), as prescribed in ASTM C522-03. S1 and S2 signify sample 1 and sample 2 of each compression ratio.

**Table 2.** Comparison of the measured static airflow resistance and resistivity of uncompressed and compressed Al-foam samples. All samples are 12.7 mm thick.

| Compression Ratio | Sample Label | Airflow Resistance (Rayls MKS) | Static Airflow Resistivity (Rayls MKS/m) |
|---|---|---|---|
| 1 | A1 (S1) | 4.4087 | 347.1433 |
| 1 | A1 (S2) | 4.0206 | 316.583 |
| 2 | B1 (S1) | 12.6603 | 996.8772 |
| 2 | B1 (S2) | 12.8243 | 1009.793 |
| 3 | C1 (S1) | 33.7666 | 2658.793 |
| 3 | C1 (S2) | 35.9671 | 2832.058 |
| 4 | D1 (S1) | 88.1775 | 6943.113 |
| 4 | D1 (S2) | 91.5772 | 7210.81 |

### 3.2. Experimental results: Uniform Al-foam configurations

#### 3.2.1. Effect of stacking on acoustic properties

In this study, we achieve a stepwise property gradient by stacking individual 12.7 mm thick disks to form the final 50.8 mm thick sample, as opposed to fabricating a single continuous sample with stepwise gradients within it. To quantify any differences occurring between stacked and continuous samples, we compare the normal incidence acoustic properties of individual 50.8 mm thick compressed and uncompressed Al-foams with uniform through-thickness properties to those measured for equivalent samples created by stacking four 12.7 mm thick disks. Fig. 8 compares the sound absorption and surface impedance properties of samples with a compression ratio of 2.

Similar results are obtained for all samples, including the uncompressed samples, and are not shown here in the interest of brevity.

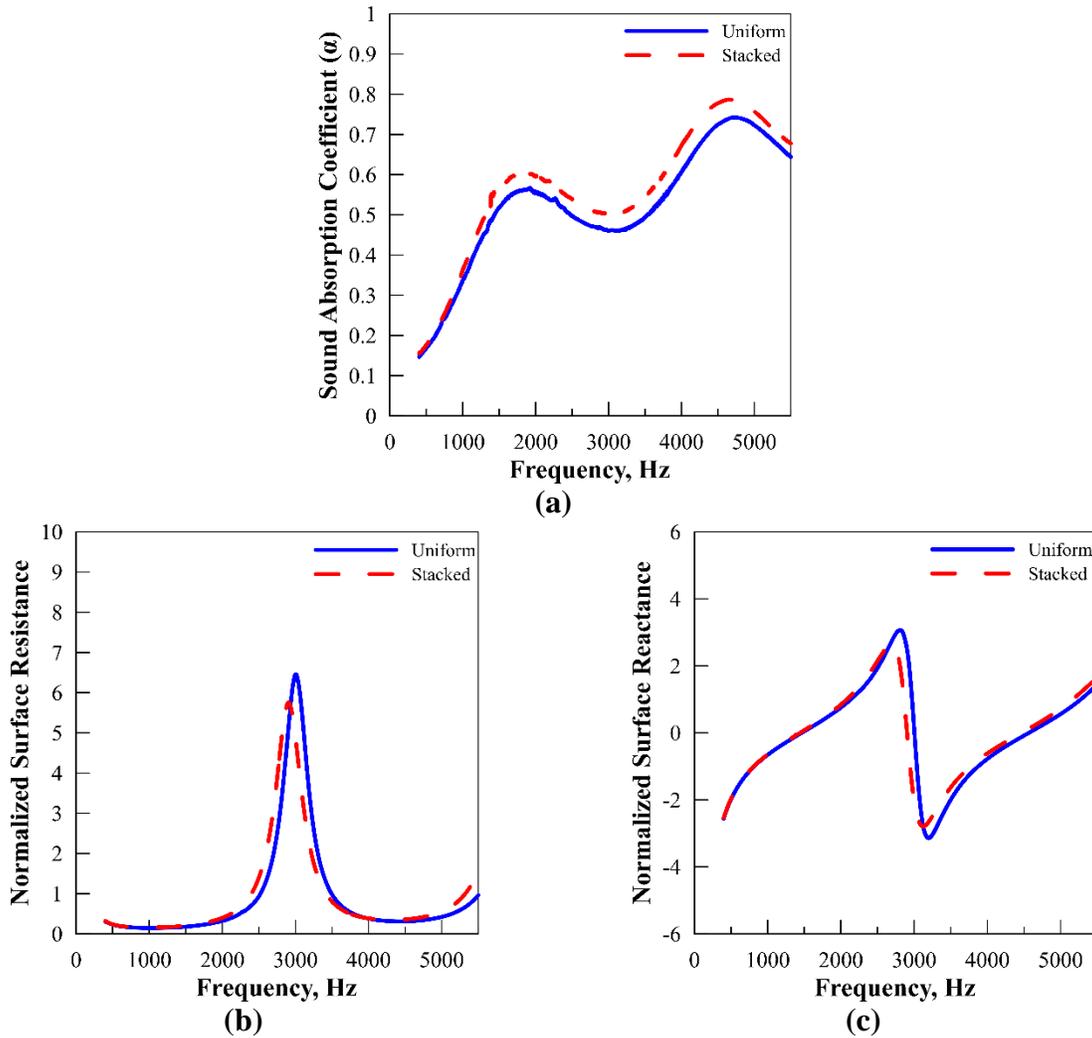

**Fig. 8.** Normal incidence **(a)** sound absorption coefficient, **(b)** normalized surface resistance, and **(c)** normalized surface reactance of Al-foam sample with a compression ratio of 2. Uniform signifies a single sample of 50.8 mm thickness, while stacked signifies an equivalent 50.8 mm thick sample created by stacking individual 12.7 mm thick disks.

While the overall trends for both samples are similar, the disk stacking results in a noticeable increase in the absorption at frequencies above 1000 Hz. The overlapping absorption behavior at lower frequencies indicates that the sample stacking does not result in an appreciable increase in the static airflow resistance. This is borne out by the overlapping low frequency asymptotes of the normalized surface resistance curves. Instead, the increase in absorption is accompanied by a corresponding shift in the absorption and impedance curve peaks to a lower frequency, indicating that the disk stacking results in a slight increase in the effective tortuosity of the sample. Since the same behavior is observed for the uncompressed samples and no core-shell differences occur in the compressed samples, this increase in tortuosity may be attributed to the change in propagation path length occurring because of the imperfect structural continuity at the layer interfaces.

*3.2.2. Effect of compression on acoustic properties*

To study the effect of compression, the acoustic properties of the uncompressed and compressed samples are measured by stacking four 12.7 mm disks within the sample holder. Fig. 9 compares the measured normal incidence sound absorption coefficient, transmission loss, and surface impedances. Overall, the absorption and transmission loss curves resemble those commonly observed for open-celled bulk materials and the acoustic energy losses are primarily resistance driven. An increase in the compression ratio increases the absorption and transmission loss over the entire frequency range. As the compression ratio increases, the absorption peaks shift to lower frequencies, though this shift is not as significant as the overall increase in the absorption values. These peak frequencies coincide with the zero-crossing of the reactance curves, showing that the absorption peaks are primarily because of depth resonances. Interestingly, while the improvements achieved in the absorption reduce with increasing compression ratios—the absorption jump is significantly greater between uncompressed and sample with compression ratio of 2 than between samples with compression ratios 3 and 4—the overall transmission loss shows increasing improvements with increased compression ratios. The improved absorption and transmission loss behavior result from the reduction in the effective pore sizes; as the compression ratio increases, the pore sizes decrease, which increases the static airflow resistance and tortuosity of the samples and results in higher thermoviscous energy losses. While the increase in airflow resistance is initially beneficial, the reduced improvements with increasingly higher compression ratios indicates that eventually the reduction in pore sizes starts resulting in excessively high airflow resistance, causing higher reflection—hence higher transmission loss—and less absorption of the incidence sound energy. While compression ratios beyond 4 were not studied here, we postulate that increasing the compression ratio beyond 4 will not result in appreciable performance benefits for sound absorption applications. Also recall that the increased compression ratios are accompanied by a corresponding increase in relative densities—a sample with compression ratio of 4 is approximately 4.6 times heavier than the uncompressed sample.

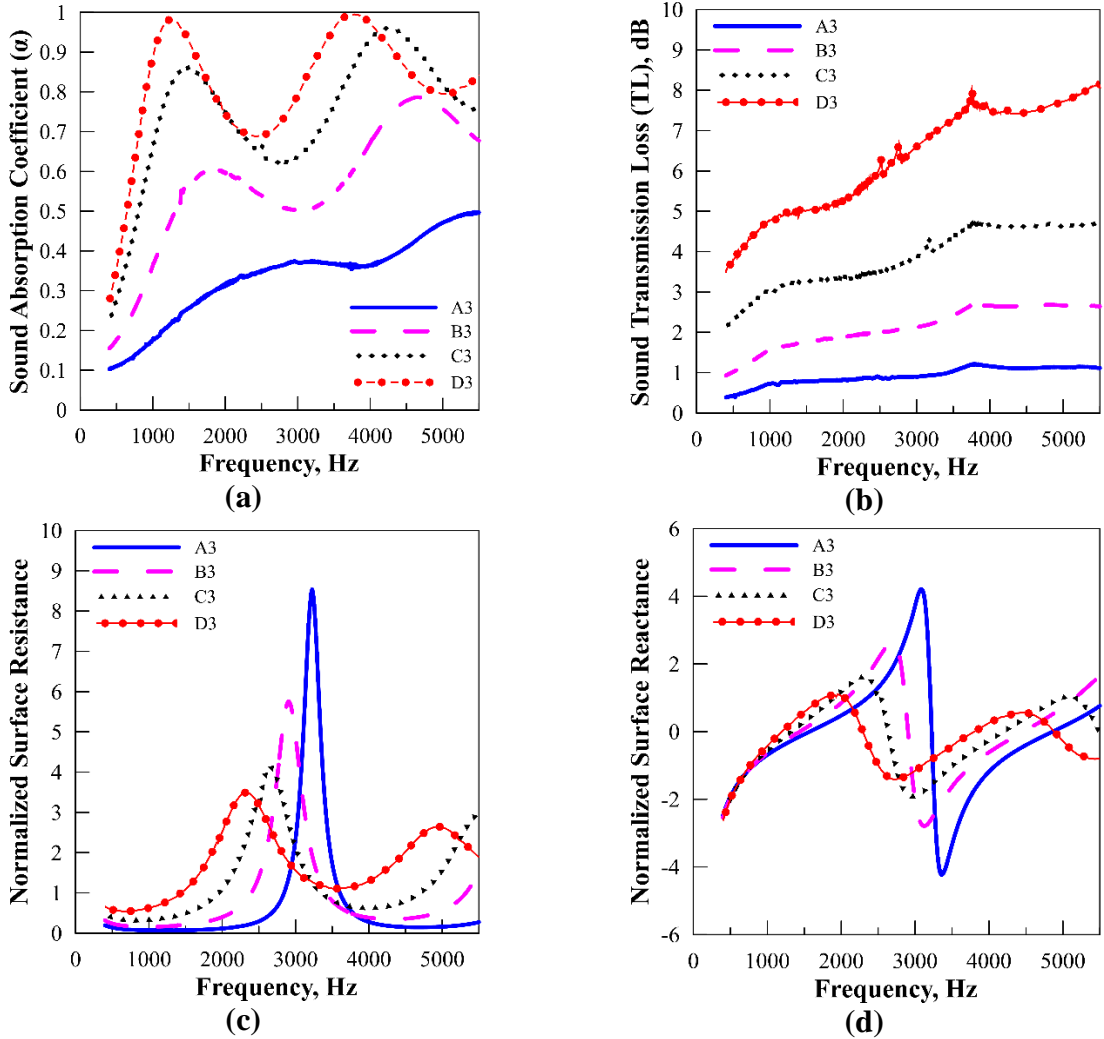

**Fig. 9.** Normal incidence **(a)** sound absorption coefficient, **(b)** sound transmission loss, **(c)** normalized surface resistance, and **(d)** normalized surface reactance of compressed Al-foam samples with 50.8 mm thickness

## 3.3. *Experimental results: Stepwise gradient Al-foams*

Increasing the compression ratio improves the acoustic absorption and transmission loss properties of open-celled Al-foams; however, this improvement also increases the flow resistance and system weight. Gradient foams provide an alternate avenue for improving these properties while incurring lower resistance and weight penalties. Here, we study the possibility of improving the acoustical performance of Al-foams by layering individual foams with different compression ratios to create stepwise property gradients. The total sample thickness for all cases is maintained at 50.8 mm. We investigate three different stepwise gradient configurations: 1-step gradients created using two Al-foams with different compression ratios; 2-step gradients created using three different compression ratios; and a 3-step gradient created using all four different compression ratios. For each gradient type, we also study the effect of sample sequence—i.e. progressively increasing or decreasing compression ratios with respect to the incident wave direction. For convenience, weights of all sample configuration are summarized in Table 3.

**Table 3.** Comparison of the total weights of various stepwise gradient configurations studied here.

| Number of Steps | Sample Label | Configuration Weight (g) | Number of Steps | Sample Label | Configuration Weight (g) |
|---|---|---|---|---|---|
| 1 | A2_B2 | 11.531 | 2 | A1_B1_C2 | 17.733 |
| 1 | A2_C2 | 15.419 | 2 | A1_B1_D2 | 21.701 |
| 1 | A2_D2 | 19.388 | 2 | A1_C1_D2 | 23.645 |
| 1 | B2_C2 | 20.046 | 2 | B1_C1_D2 | 25.959 |
| 1 | B2_D2 | 24.014 | 3 | A1_B1_C1_D1 | 19.717 |
| 1 | C2_D2 | 27.903 | | | |

*3.3.1. 1-step configurations*

Fig. 10. shows the measured absorption and transmission loss curves for 1-step configurations, where each uniform step of 25.4 mm thickness is created using two 12.7 mm samples with the same compression ratios. As compared to the uncompressed sample, shown in Fig. 9, the absorption and transmission loss of all stepwise configurations show a marked improvement. As expected, for samples with one uncompressed layer, the improvement is greater when the second layer has a greater compression ratio. Overall, the highest absorption and transmission loss are obtained for samples composed using foams with compression ratios 3 and 4 (C2_D2 and D2_C2). The sample with increasing compression ratio with respect to the incident wave direction, C2_D2, achieves total sound absorption at 4100 Hz, while the decreasing compression ratio sample, D2_C2, achieves total absorption at 3800 Hz. Comparing equivalent configurations with increasing and decreasing compression ratios, we observe that all configurations with decreasing compression ratios show two high absorption peaks separated by a reduced absorption trough region. Conversely, configurations with increasing compression ratios exhibit comparatively lower absorption peaks; however, the absorption in the trough (or mid-peak) region is significantly improved and higher absorption is maintained over a wider frequency band. This behavior is more exaggerated for configurations with layers of greater property contrasts, such as the A2_C2, A2_D2, and B2_D2 samples. Further, configurations with stepwise increasing compression ratios show poorer absorption performance at lower frequencies. This indicates that the low frequency sound absorption behavior of layered configurations is primarily dictated by the airflow resistivity of the incident layer. Overall, the potential of improving acoustic behavior using graded configurations is reinforced by observing that the D2_A2 provides comparable absorption and transmission loss to a uniform *D* sample while being 40% lighter.

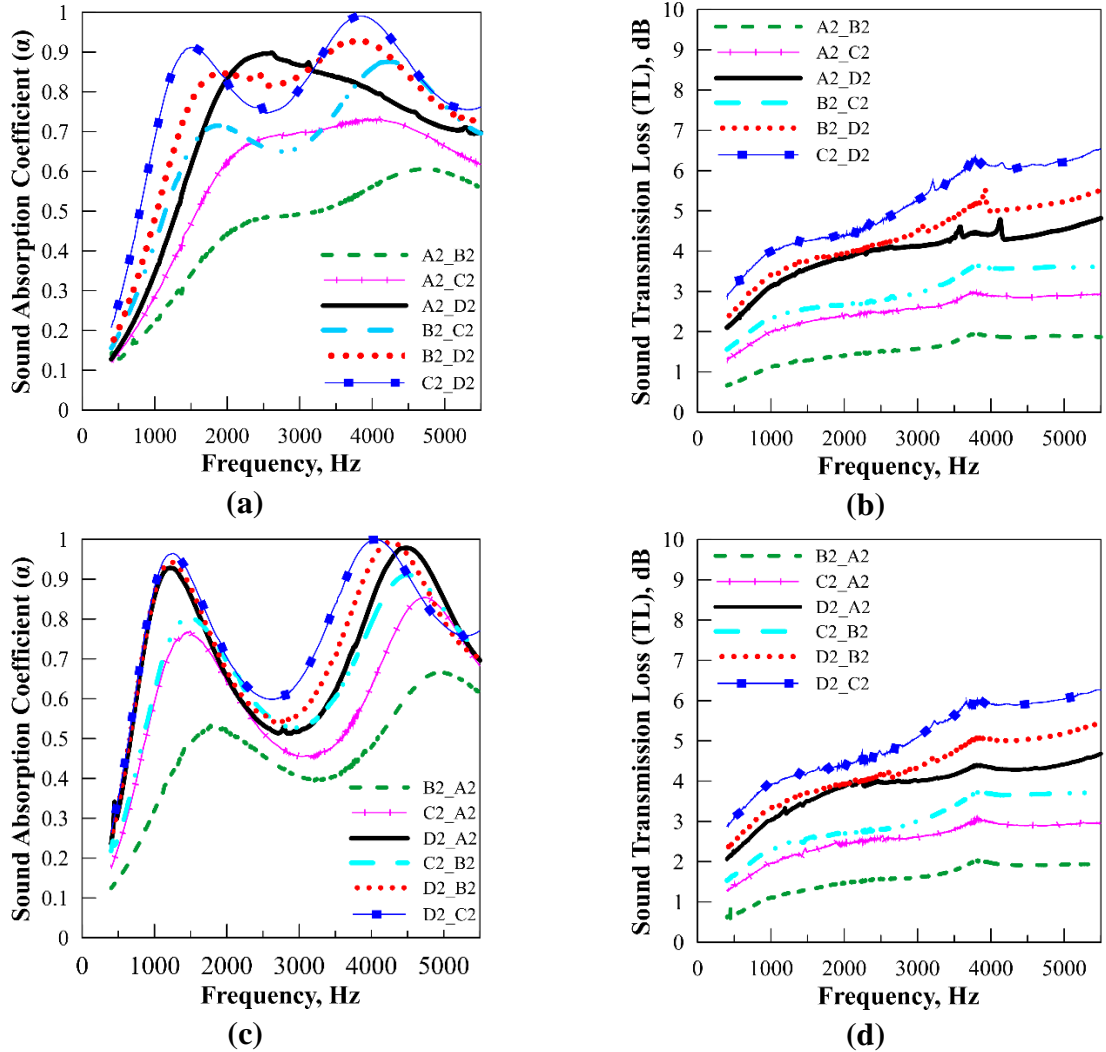

**Fig. 10.** Normal incidence **(a, c)** sound absorption coefficient, and **(b, d)** sound transmission loss of 1-step configurations with **(a, b)** increasing compression ratios and **(c, d)** decreasing compression ratios, respectively. The corresponding impedance curves are provided in Appendix A.

### 3.3.2. 2- and 3-step configurations

Fig. 11 shows the absorption and transmission loss measured for various 2-step configurations and the 3-step configuration. The trends are like those observed for the 1-step configurations: decreasing compression ratios result in higher absorption peaks separated by a low absorption trough; increasing compression ratios result in lower absorption peaks occurring at comparatively higher frequencies but provide a flatter absorption response over a wider frequency range. Samples with greater inter-layer compression ratio contrasts provide better absorption characteristics. Note that all 2-step and 3-step configurations shown here are lighter than the uniform D sample and the C2_D2 configuration. Further, while the A1_B1_C1_D1 samples seems to provide absorption behavior inferior to the 2-step configurations other than the A1_B1_C2 sample, the total weight of this configuration is comparable to the D2_A2 configuration and is also lower than the outperforming 2-step configurations. Thus, graded configurations provide a significant scope for tailoring the acoustical performance of Al-foams while minimizing the resultant weight penalties.

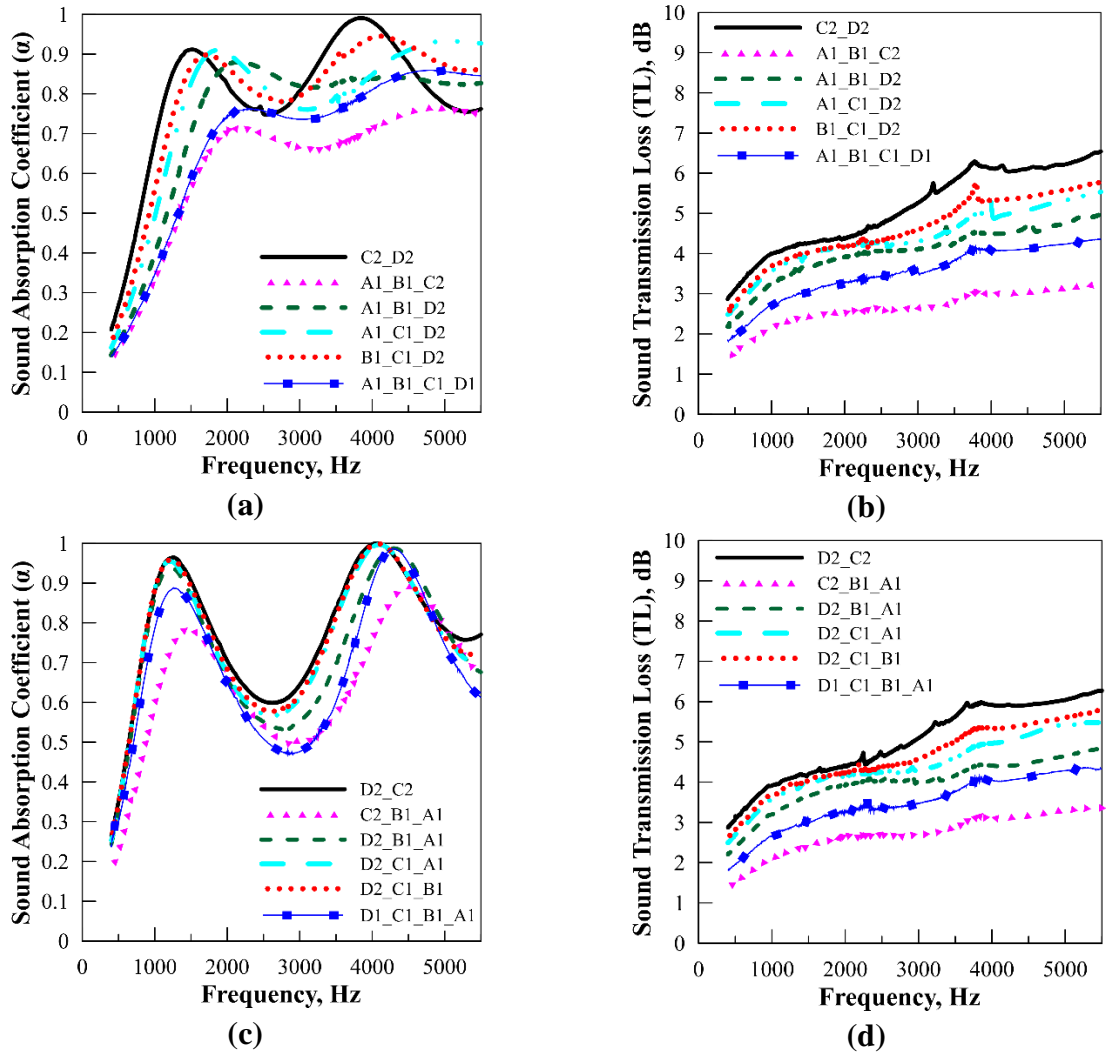

**Fig. 11.** Normal incidence **(a, c)** sound absorption coefficient, and **(b, d)** sound transmission loss of 2-step and 3-step configurations with **(a, b)** increasing compression ratios and **(c, d)** decreasing compression ratios, respectively. The corresponding impedance curves are provided in Appendix B.

### 3.4. *Transfer Matrix Modeling*

The experimental results show that Al-foams with stepwise property gradients can provide superior acoustical performance as compared to uniform Al-foams; however, the experiments are limited to layer thicknesses of 12.7 mm and its integer multiples. Here, we use the transfer matrix modeling approach to overcome this limitation and provide a feasible route towards future optimization studies. For this approach, we first measure the thickness-independent characteristic impedance and wave numbers using uniform 25.4 mm thick Al-foam samples. We then use these properties to formulate the thickness-dependent transfer matrices for each layer, given by Eq. (8), and finally assemble them into the global transfer matrix, given by Eq. (9). The global transfer matrix is then used to predict a chosen configuration's surface impedance, absorption coefficient, and transmission loss behavior.

### 3.4.1. Model verification

To verify the method, we compare the transfer matrix modeling predictions with the experimental measurements obtained previously. Fig. 12 compares the predicted and experimental measurements obtained for the uncompressed sample of 50.8 mm thickness. As before, the experimental results are obtained by constructing the 50.8 mm samples using individual 12.7 mm thick layers. The predictions are calculated by assuming the sample is constructed using either 2, 4, or 8 individual layers. Experimental results from both the four-microphone and two-microphone impedance tube methods are shown for completeness. Note that we calculate the absorption coefficient by assuming a rigidly terminated sample.

Overall, the predicted and measured surface impedance curves show an excellent match over the entire frequency range. As expected, the predictions calculated using various layer thicknesses overlap each other. The absorption and transmission loss curves show a good match until approximately 3800 Hz; beyond this frequency, though the trends remain consistent with the experiments, the transfer matrix method over-predicts both the properties. Further investigation shows that these deviations correspond to the measurement deviations occurring in the characteristic properties measured using the four-microphone method. Theoretically, for uniform isotropic samples, the four-microphone approach provides thickness-independent characteristic properties; however, in practice, the accuracy of these measurements is sensitive to the sample thickness and measurement noise. Fig. 13 compares the characteristic properties measured for the uncompressed Al-foam using test samples of 12.7, 25.4, and 50.8 mm thicknesses. Although the characteristic impedances and the wave numbers (i.e., the real part of the wave number) for all thicknesses are quite close, the attenuation constants (i.e., the imaginary part of the wave number) show a dependence on the sample thickness. The deviations occurring above 3800 Hz in the attenuation constant measured for the 25.4 mm thick sample directly correspond to the over-predictions in the absorption and transmission loss properties. These measurement issues also contribute to the overall 'wavy' characteristics of the calculated properties. For brevity, only the comparisons for the uncompressed sample are shown here; all other samples showed closer match between the experiments and predictions because of their higher absorption and transmission loss values, which result in less noisy four-microphone data.

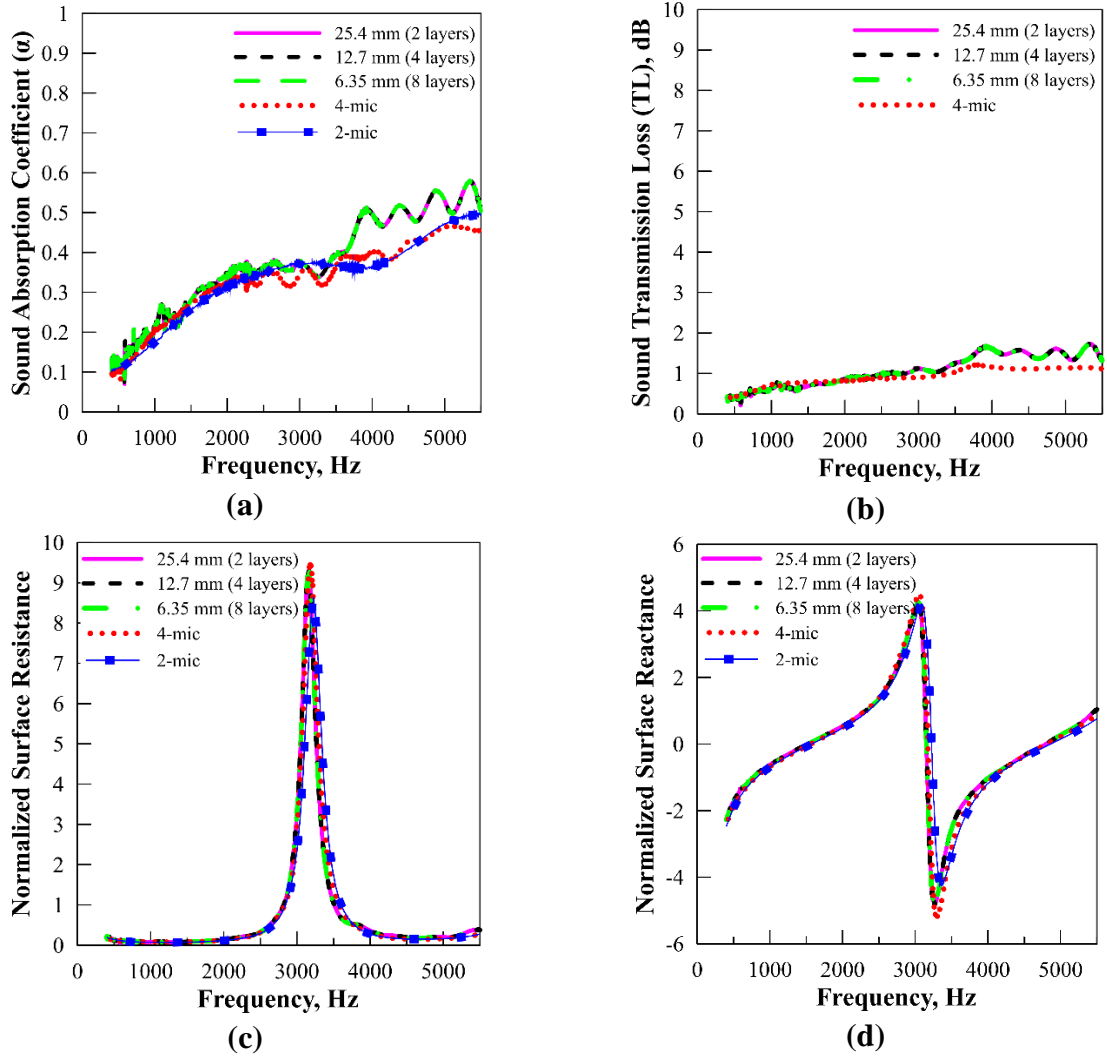

**Fig. 12.** **(a)** Normal incidence sound absorption coefficient, **(b)** normal incidence sound transmission loss, **(c)** normalized surface resistance, and **(d)** normalized surface reactance using multi-layered transfer matrix modeling for *A* with 50.8 mm thickness.

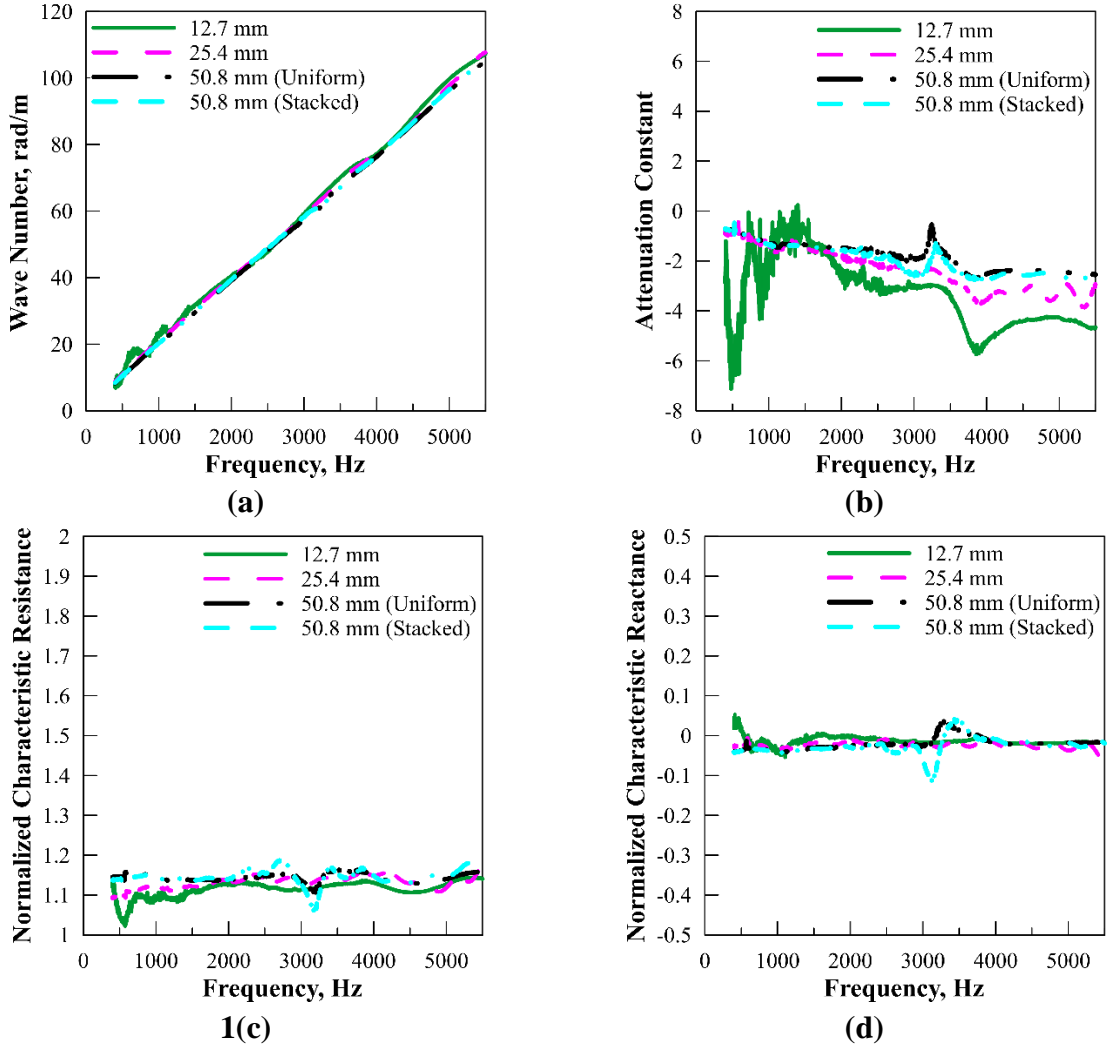

**Fig. 13.** **(a)** Wave number, **(b)** attenuation constant, **(c)** normalized characteristic resistance, and **(d)** normal characteristic reactance measured using four-microphone impedance tube method for an uncompressed sample with 12.7, 25.4, and 50.8 mm thicknesses.

Further model verification is provided by comparing the predictions for the stepwise property gradient configurations with their experimental measurements. For brevity, only comparisons for single 1-step (B2_D2), 2-step (B1_C1_D2), and 3-step configuration (A1_B1_C1_D1) are presented in Fig. 14.

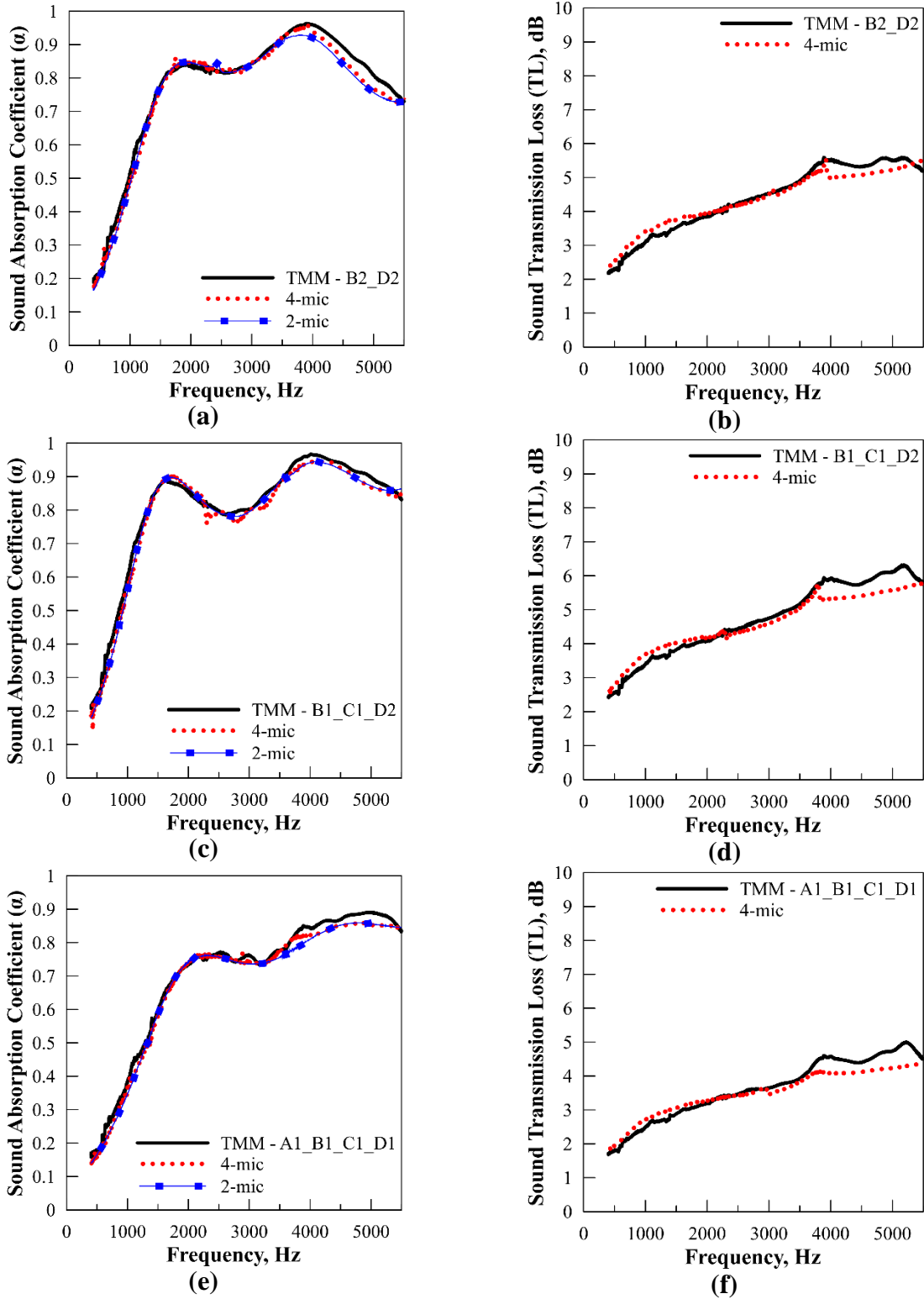

**Fig. 14.** Comparison of the predicted and measured normal incidence sound absorption coefficient and transmission loss for **(a, b)** 1-step sample (B2_D2), **(c, d)** 2-step sample (B1_C1_D2), and **(e, f)** 3-step sample (A1_B1_C1_D1), respectively. The corresponding impedance curves are provided in Appendix C.

### *3.4.2. Parametric studies*

Experimental results from section 3.3 show that Al-foams with stepwise relative density gradients can provide improved acoustical performance without a significant weight penalty. However, the experimental studies were limited to individual step thicknesses of 12.7 mm. Here, we leverage the transfer matrix model to overcome this thickness limitation and conduct studies to identify stepwise gradient configurations that provide optimal performance-to-weight ratios. Our primary target is to achieve an improved acoustic absorption performance between 1000–5000 Hz—the typical frequency range of interest for aerospace applications. Experimentally, the 50.8 mm thick sample with compression ratio of 4 (D3) provides the best acoustic absorption; it is also the heaviest configuration. However, though this configuration provides two absorption peaks with almost total sound absorption (1250 Hz and 3784 Hz), these peaks are separated by a trough where the absorption coefficient reduces to 0.69. Here, we attempt to identify the lightest stepwise gradient configuration that retains the high absorption peaks achieved by D3 while improving the absorption performance in the frequency range between the peaks. The varied parameters are the layer thickness and the stacking sequence; we maintain the total configuration thickness at 50.8 mm for all cases. We allow the individual layer thickness to vary from 1 mm up to 50.8 mm and limit our study to configurations with a maximum of 6 steps (i.e., 7 individual layers). Since our target is to identify the configurations with the best absorption-to-weight ratios, we also track the total weight of each configuration as an additional variable of interest. The obtained results are summarized in Fig. 15 and Table 4. For brevity, we only show the results obtained from 5 configurations that were identified to provide the best performance. Though the configurations are not optimized for transmission loss performance, we provide the predicted transmission loss of each configuration for completeness. The absorption and transmission loss curves—measured using the two- and four-microphone methods, respectively—for the D3 sample are also provided for reference. In Table 4, the numbers between the brackets identify the layer thickness in mm; the first layer in the sequence is assumed to be the incident surface of the configuration.

**Table 4.** Stacking sequences and total weights of various configurations presented in Fig. 15. The D3 configuration is used as the reference configuration for weight reduction calculations.

| Configuration | Stacking sequence (thickness in mm) | Total weight (g) | Reduced weight (%) |
|---|---|---|---|
| Reference | D3 | 31.871 | - |
| 1 | B(6.5)_C(6.35)_D(38.1) | 28.915 | 9.275 |
| 2 | A(6.35)_B(6.35)_C(12.7)_D(25.4) | 24.801 | 22.18 |
| 3 | D(5)_A(6)_B(3.6)_C(5.6)_D(21.4)_B(3.6)_C(5.6) | 24.945 | 21.73 |
| 4 | D(5)_A(2)_B(9.2)_C(9.2)_D(25.4) | 26.604 | 16.522 |
| 5 | D(11.95)_A(1)_C(11.95)_A(1)_D(11.95)_A(1)_C(11.95) | 26.662 | 16.341 |

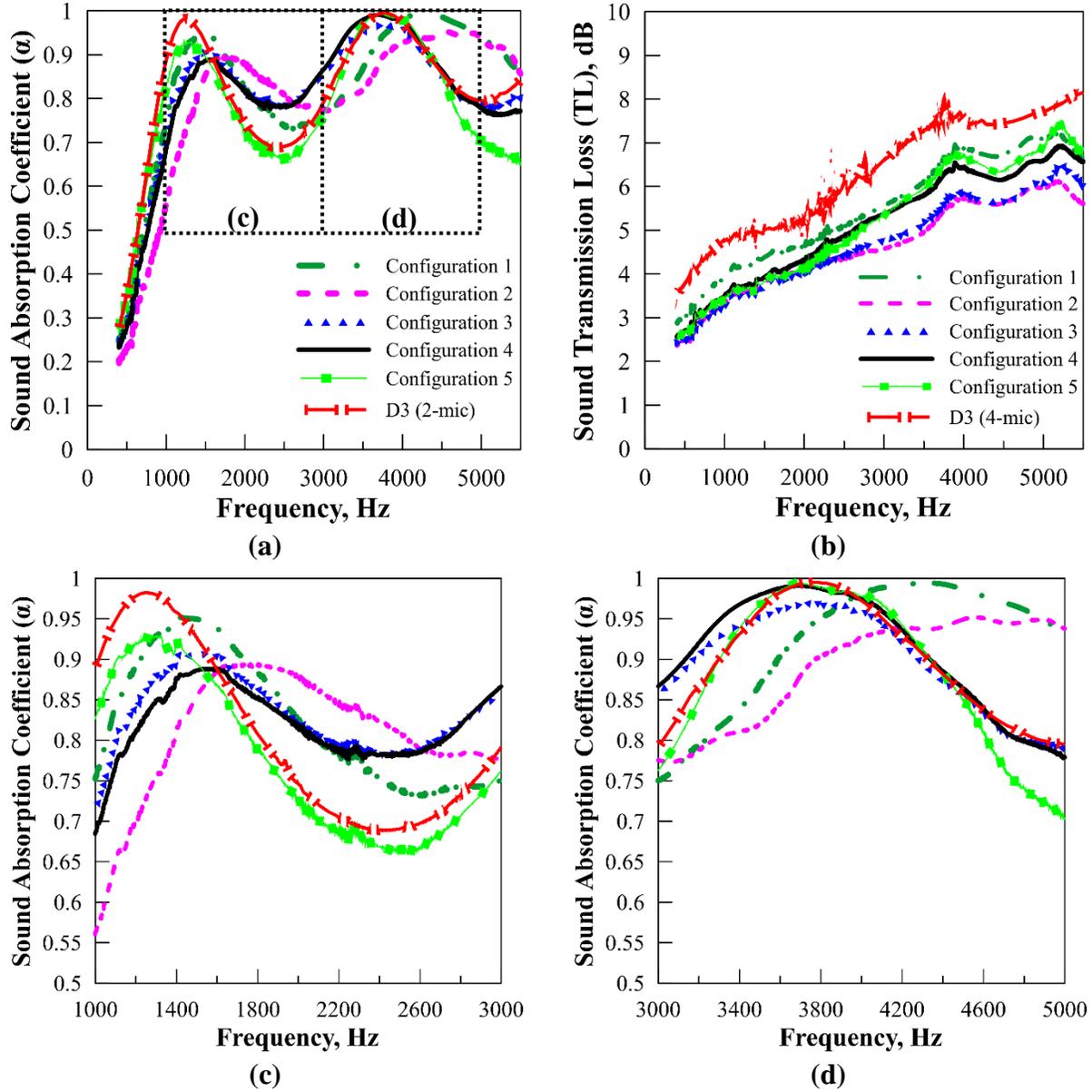

**Fig. 15.** Comparison of **(a)** normal incidence sound absorption coefficient, **(b)** normal incidence transmission loss for the configurations shown in Table 4**.** using multi-layered transfer matrix modeling; **(c)** and **(d)** show close-ups of the sound absorption coefficient for the frequency range 1000-3000 Hz and 3000-5000 Hz respectively.

Configurations 1 and 2 are created by stacking the Al-foams in a stepwise increasing relative density (or compression ratio) sequence. In both configurations, the sample with a compression ratio of 4 (D) forms the thickest layer and is placed furthest away from the incident side. Configuration 1, with sample D forming a 38.1 mm thick layer and making it the heaviest configuration, achieves absorption peak magnitudes closest to those achieved by the D3 configuration; however, addition of the preceding layers with lower relative densities helps improve the mid-peak absorption to 0.73. The first absorption peak shifts from 1250 Hz to 1416 Hz and reduces from 0.98 to 0.95. This is in accordance with the previous observation that an increasing relative density configuration results in lower absorption peak magnitudes but better mid-peak performance. Addition of the A layer in configuration 2, accompanied by an increased thickness of layer C and decreased thickness of layer D, further reduces the first absorption peak to 0.89 and shifts it to 1826 Hz. However, the increasing gradient configuration helps improve the mid-peak absorption further—this configuration provides the best performance between 1600 and 3100 Hz and provides a minimum absorption coefficient of 0.77 over the entire mid-peak frequency range. It is also the lightest configuration, providing 22.18% weight reduction as compared to the reference D3 configuration. Configuration 3 provides the next best weight reduction (21.73%) by splitting the total thickness into 7 layers. This configuration reduces the first absorption peak to 0.9, achieves mid-peak absorption above 0.78 over the entire frequency range, and closely follows the performance of the D3 configuration above 3500 Hz. Its performance is closely mirrored by configuration 4, which is created by sandwiching an increasing relative density sequence of A, B, and C samples between two D sample layers. Finally, configuration 5 provides an example of a stepwise gradient sample that achieves a performance similar to the D3 sample over the 400–4500 Hz frequency range, while being 16.34% lighter than it. For all the cases studied, we noticed that configurations with the highest relative density sample forming the first layer provide the highest absorption peak magnitudes, while addition of sequences with increasing relative densities helps improve the mid-peak absorption performance. Thus, stepwise relative density gradient configurations allow the design of lightweight sound absorption packages with performance tailored for the specific noise reduction application.

**Conclusions**

We present the normal incidence acoustical properties of uncompressed and compressed open-celled Al-foams. The effect of compression on the cellular microstructure, static airflow resistivity, and acoustic absorption and transmission behavior is studied. Using image analysis, we show that compressing the Al-foams distorts the struts and reduces the Al-foam's effective open porosity. A compression ratio of 4 causes densification and some amount of strut breakage. This cellular distortion increases the airflow resistivity and relative density of the Al-foam.

The acoustical properties under plane wave, normal incidence conditions are studied using two- and four-microphone impedance tube setups. It is shown that the compression driven cellular distortion helps improve the absorption and transmission loss performance, with the most compressed Al-foams providing the best acoustical performance. The measured impedances show that the absorption is primarily due to resistive effects with two distinct absorption peaks occurring due to thickness-resonances under the rigid termination test condition. The acoustic behavior of stepwise relative density gradient foams is then studied by stacking individual disks with different compression ratios within the sample holder. It is seen that the stepwise gradient improves the overall performance of the samples. The effect of different stepwise relative density gradient configurations—either increasing or decreasing with respect to the incident wave direction—is studied. We show that both configurations result in distinct changes in the absorption behavior; the transmission loss behavior remains unchanged. Configurations with stepwise increasing relative density gradients result in comparatively lower absorption peaks that occur at higher frequencies, as compared to the configurations with stepwise decreasing relative density gradients. However, the increasing gradients provide significantly better absorption performance between the two peaks as compared to the decreasing gradients. Further, stronger interlayer property contrasts result in greater absorption improvements.

Finally, we present an experimentally informed transfer matrix method to predict the acoustical properties of various sample depths and relative density gradients. We validate the method using experimental results and then use it to conduct parametric studies aimed at achieving high sound absorption performance over the 1000-5000 Hz frequency range while reducing the overall weight of the sample. We present five different sample configurations with different sound absorption performance and total weights—all configurations provide excellent absorption-to-weight performance as compared to an equivalent thickness configuration that provides the best absorption performance. These results show that stepwise relative density open-celled Al-foams provide a feasible route to creating performance-tailored, multifunctional noise reduction packages for high temperature and pressure environments where the use of traditional sound absorption materials is infeasible.

**Acknowledgement**

This work was supported by a NASA EPSCoR Cooperative Agreement Notice (Grant Number: 80NSSC19M0153). The authors would like to thank Mike Jones, Brian Howerton, Noah Schiller, and Daniel Sutliff for their helpful comments and suggestions. The contributions of Brittany Wojciechowski and Jake Puppo are gratefully acknowledged for their role in data acquisition and sample fabrication.

**Appendix A**

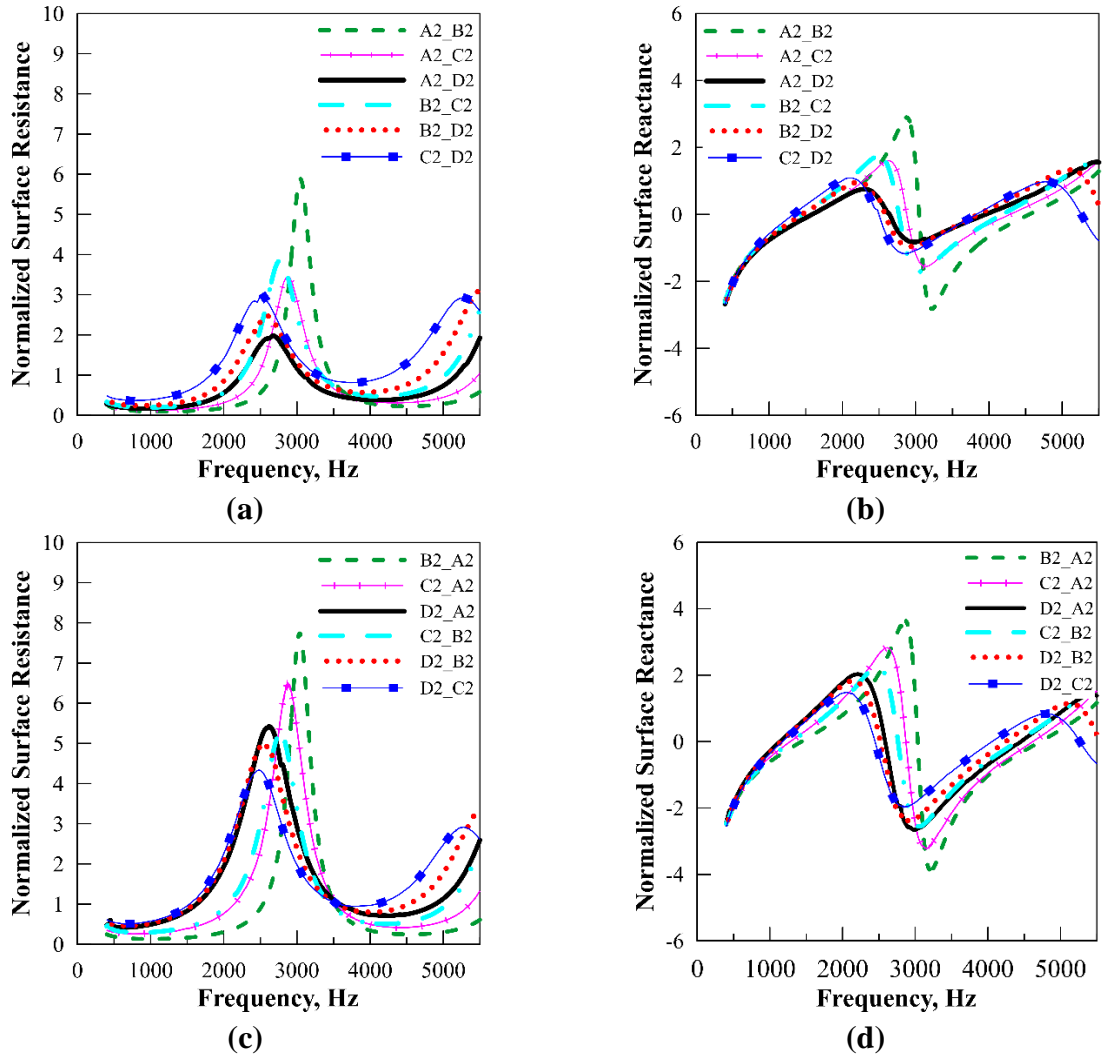

**Fig. A.1.** **(a, c)** Normalized surface resistance, and **(b, d)** normalized surface reactance of 1-step configurations shown in Fig. 10, with **(a, b)** increasing compression ratios and **(c, d)** decreasing compression ratios, respectively.

**Appendix B**

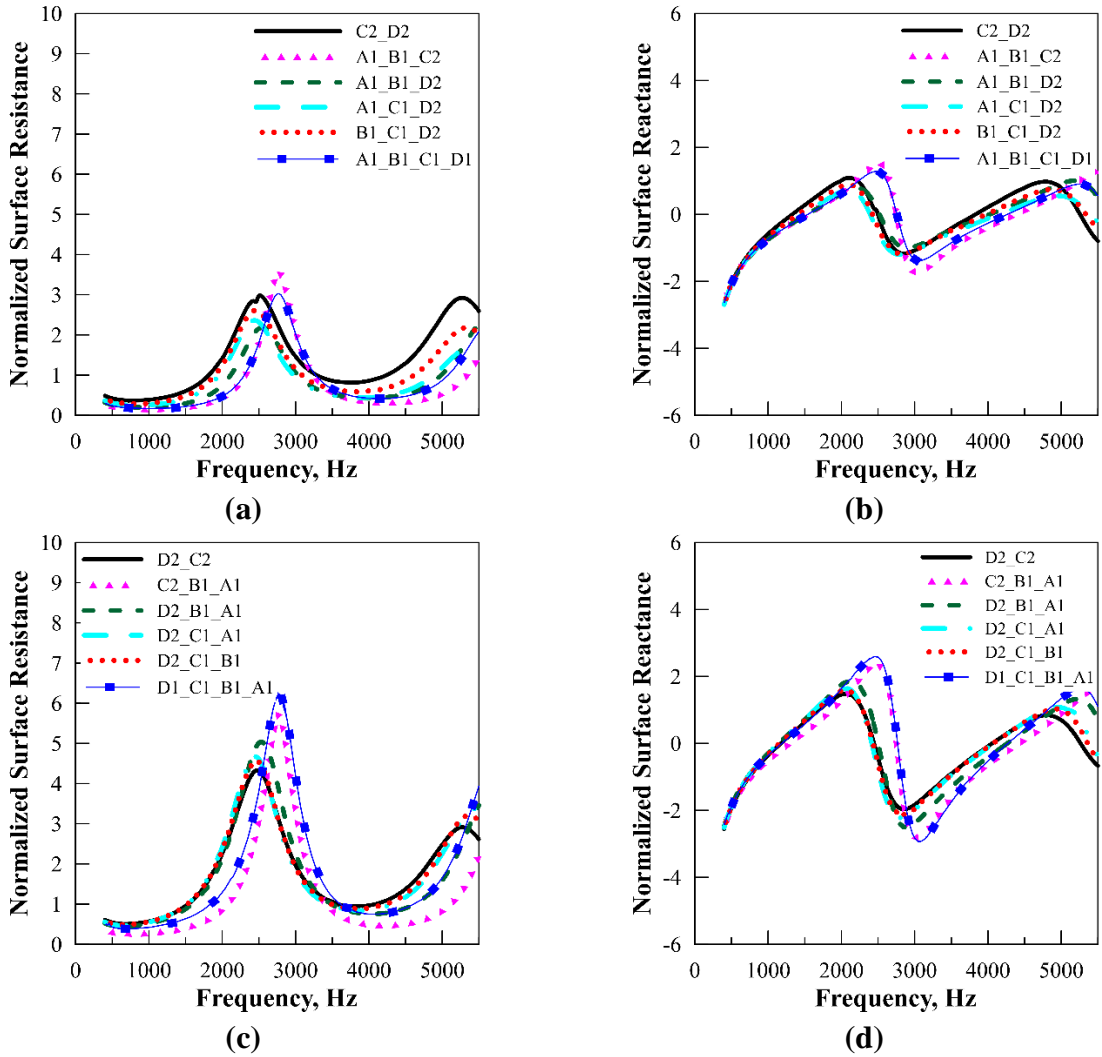

**Fig. B.1. (a, c)** Normalized surface resistance, and **(b, d)** normalized surface reactance of 2-step and 3-step configurations shown in Fig. 11, with **(a, b)** increasing compression ratios and **(c, d)** decreasing compression ratios, respectively.

**Appendix C**

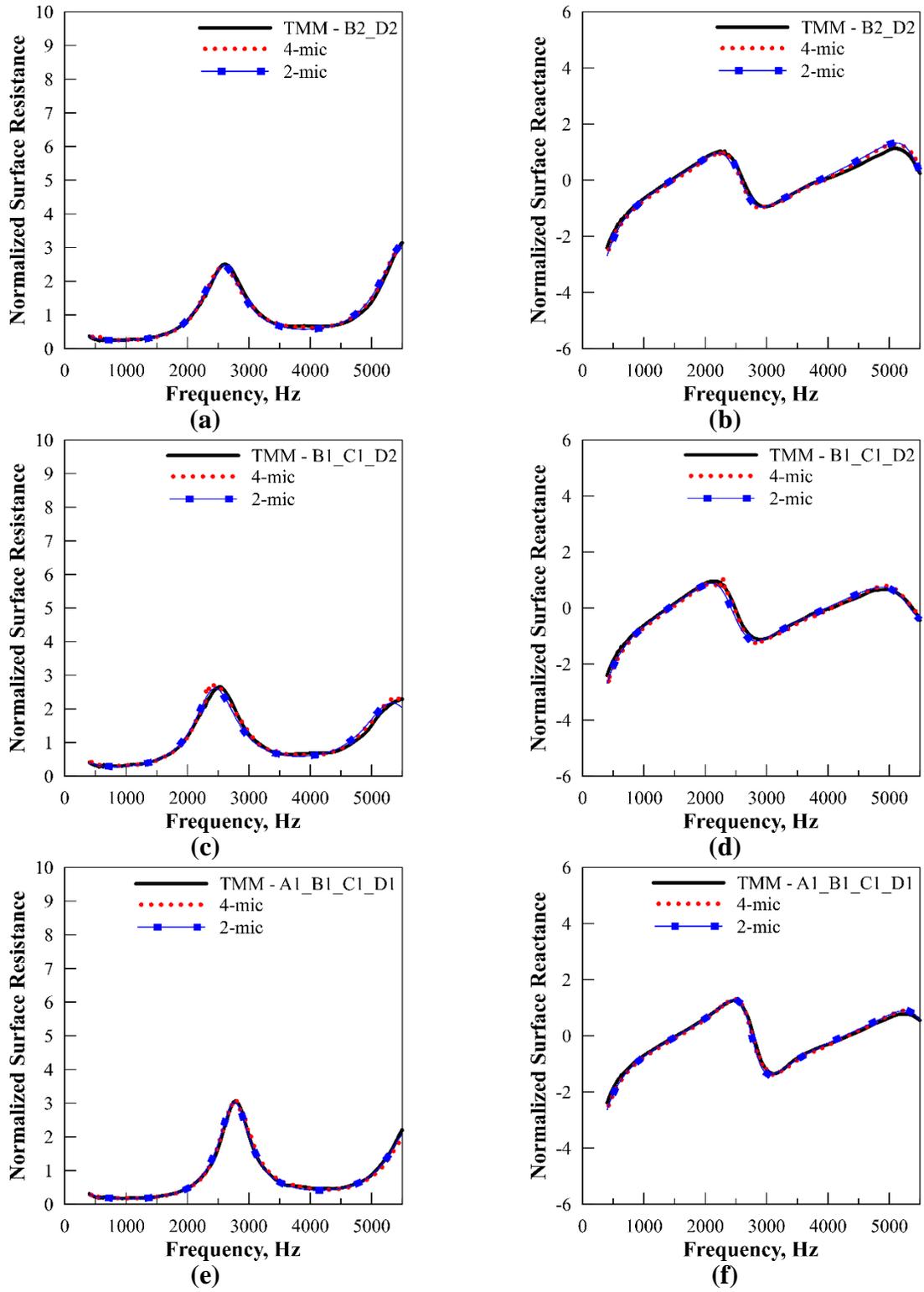

**Fig. C.1.** Comparison of the predicted and measured normalized surface resistance and normalized surface reactance for configurations shown in Fig. 14. **(a, b)** 1-step sample (B2_D2), **(c, d)** 2-step sample (B1_C1_D2), and **(e, f)** 3-step sample (A1_B1_C1_D1), respectively.

**Appendix D**

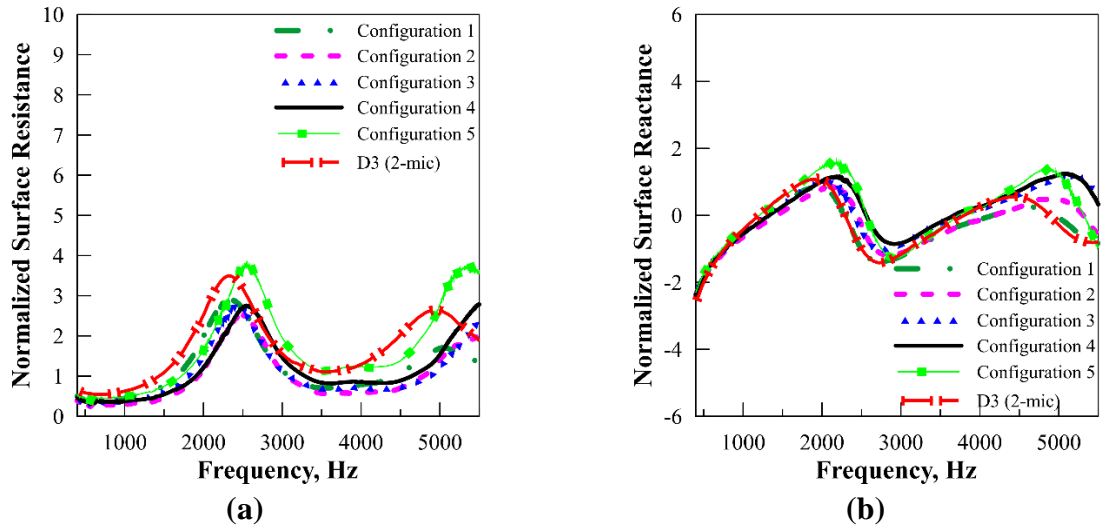

**Fig. D.1.** Comparison of **(a)** normalized surface resistance, **(b)** normalized surface reactance for the configurations shown in **Table 4.** using multi-layered transfer matrix modeling